\documentclass[aps,preprint,groupedaddress]{revtex4}
\usepackage{amsfonts}
\usepackage{mathrsfs}
\usepackage{graphicx}
\usepackage{amsmath}
\usepackage{amssymb}
\usepackage{amsbsy}
\usepackage{bm}
\usepackage{soul}
\usepackage{xcolor}
\newcommand{\blue}{{}}
\makeatletter

\newcommand{\Rmnum}[1]{\expandafter\@slowromancap\romannumeral #1@}
\makeatother

\begin{document}
    
\title{Eight challenges in developing theory of intelligence}
\author{Haiping Huang}
\email{huanghp7@mail.sysu.edu.cn}
\affiliation{PMI Lab, School of Physics,
Sun Yat-sen University, Guangzhou 510275, People's Republic of China}

\date{\today}
\begin{abstract}
   A good theory of mathematical beauty is more practical than any current observation, as new predictions about physical reality can be self-consistently verified. This belief applies to the current status of understanding deep neural networks including large language models and even the biological intelligence. Toy models provide a metaphor of physical reality, allowing mathematically formulating that reality (i.e., the so-called theory), which can be updated as more conjectures are justified or refuted. One does not need to pack all details into a model, but rather, more abstract models are constructed, as complex systems like brains or deep networks have many sloppy dimensions but much less stiff dimensions that strongly impact macroscopic observables. This kind of bottom-up mechanistic modeling is still promising in the modern era of understanding the natural or artificial intelligence. Here, we shed light on eight challenges in developing theory of intelligence following this theoretical paradigm. \blue{Theses challenges are representation learning, generalization, adversarial robustness, continual learning, causal learning, internal model of the brain, next-token prediction, and finally the mechanics of subjective experience.}
\end{abstract}

\maketitle
\section{Introduction}
Brain is one of the most challenging subjects to understand. \blue{The brain is complex with many levels of temporal and spatial complexity~\cite{ND-2014}, allowing for coarse-grained descriptions at different levels, especially in theoretical studies}. More abstract models lose the ability to generate predictions on low level details, but bring the conceptual benefits of explaining precisely how the system works, and the mathematical description may be universal, independent of details (or sloppy variables)~\cite{JNS-2023}. One seminal example is the Hopfield model~\cite{Hopfield-1982}, where the mechanism underlying the associative memory observed in the brain was precisely isolated~\cite{Amit-1987,Amit-1993}. There is a resurgence of research interests in Hopfield networks in recent years due to the large language models~\cite{Hopfield-2020,Krotov-2020}. 

In Marr's viewpoint~\cite{Marr-1982}, understanding a neural system can be divided into three levels: computation (which task the brain solves),
algorithms (how the brain solves the task, i.e., information processing level), and implementation (neural circuit level). \blue{Following the first two levels, researchers designed artificial neural networks to solve challenging real-world problems, such as powerful deep learning~\cite{Deep-2015a,Deep-2015b}}. However, biological details are also being incorporated into 
models of neural networks~\cite{Abbott-2016,Kording-2016,DL-2019a,BPbrain-2020}, and even used to design new learning rules~\cite{BIL-2023}. Indeed, neuroscience studies of biological mechanisms of 
perception, cognition, memory and action have already provided a variety of fruitful insights inspiring the empirical or scientific studies of artificial neural networks, which in turn
inspires the neuroscience researchers to design mechanistic models to understand the brain~\cite{Yamin-2016,Saxe-2020,Neuron-2017}. Therefore, it is promising to integrate 
physics, statistics, computer science, psychology, neuroscience and engineering to reveal inner workings of deep (biological) networks and 
even intelligence with testable predictions~\cite{Ma-2022}, \blue{rather than using a black box (e.g., deep artificial neural networks) to understand the other black box (e.g, brain or mind)}. In fact, the artificial intelligence may follow different principles from the natural intelligence, but both can inspire each other, \blue{which may lead to establishment of a coherent mathematical physics foundation for either artificial or biological intelligence}. 

\begin{figure}
    \centering
\includegraphics[width=0.8\textwidth]{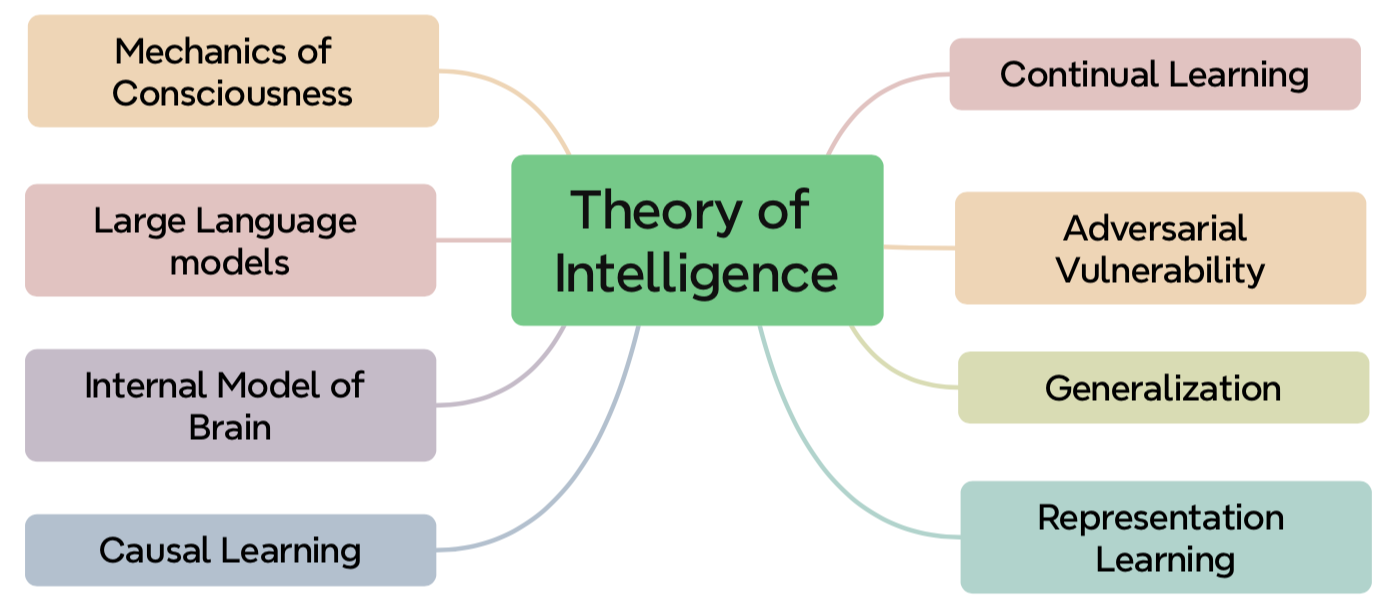}
        \caption{\blue{Schematic illustration of eight open challenging problems toward theory of intelligence.} } 
        \label{fig1}
    \end{figure}

The goal of providing a unified framework for neural computation is very challenging and even impossible. Due to re-boosted interests in neural networks, there appear
a lot of important yet unsolved scientific questions. We shall detail these challenging questions below~\footnote{Most of them were roughly provided in the book of statistical mechanics of neural networks~\cite{HH-2022}. Here we give a significantly expanded version.}, and provide our personal viewpoints toward a 
statistical mechanics theory solving these fundamental questions, based on first principles in physics. \blue{These open scientific questions toward theory of intelligence are summarized in Figure~\ref{fig1}}.

\section{Challenge \Rmnum{1}---Representation learning}
\blue{Given raw data (or input-output pairs in supervised learning), one can
ask what a good representation is and how the meaningful representation is achieved in deep neural networks}. We have not yet satisfied answers for these questions.
A promising argument is that entangled manifolds at earlier layers of a deep hierarchy are gradually disentangled into linearly separable features at output
layers~\cite{DiCarlo-2007,Bengio-2013,Brahma-2016,Huang-2018b,Cohen-2020}. This manifold separation perspective is also promising in system neuroscience studies of associative learning by
separating overlapping patterns of neural activities~\cite{Alex-2019}. However, an analytic theory of the manifold transformation is still lacking, prohibiting us from a full understanding of which key 
network parameters control the geometry of manifold, and even how learning reshapes the manifold. For example, the correlation among synapses (e.g., arising during learning) will attenuate the decorrelation process along the network depth, but encourage dimension reduction compared to their orthogonal counterparts~\cite{Huang-2020weak,Huang-2018b}. \blue{This result is derived by using mean-field approximation and coincides with empirical observations~\cite{Huang-2022}}. In addition, there may exist other biological plausible factors such as normalization, attention, homeostatic control impacting the manifold transformation~\cite{Turri-2004,Reynolds-2009}, which can be incorporated into a toy model in future to test the manifold transformation hypothesis. 

Another argument from information theoretic viewpoints demonstrates that 
the input information is maximally compressed into a hidden representation whose task-related information should be maximally retrieved at the output layers, according to the information bottleneck theory~\cite{Tishby-2017,Stefano-2017}.
In this sense, an optimal representation must be invariant to nuisance variability, and their components must be maximally independent, which may be related to causal factors (latent causes) explaining the sensory inputs (see the following fifth challenge). In a physics language, a coarse-grained (or more abstract) representation is formed in deeper layers compared to the fine-grained representation in shallower layers. How microscopic interactions among synapses determine this representation transformation remains elusive and thus deserves future studies; \blue{a few recent works started to address the clustering structure in the deep hierarchy~\cite{Li-2020,Li-2023,Barra-2023,Xie-2024}. To conclude, the bottom-up mechanistic modeling would be fruitful in dissecting mechanisms of representation transformation.}

\section{Challenge \Rmnum{2}--- Generalization} 
\blue{Studying any neural learning system must consider three ingredients: data, network and algorithm (or DNA of neural learning). The generalization ability refers to the computational
 performance that the network is able to implement the rule well in unseen examples.} Therefore, intelligence can be considered to some extent as the ability of generalization, especially given very few examples for learning.
Therefore, the generalization is also a hot topic in current studies of deep learning. Traditional statistical learning theory claims that over-fitting effects should be strong when 
the number of examples is much less than the number of parameters to learn, which thereby could not explain the current success of deep learning.
A promising perspective is to study the causal connection between the loss landscape and the generalization properties~\cite{Huang-2014pre,Baldassi-2016,Spigler-2019,Zou-2020}.
For a single layered perceptron, a statistical mechanics theory can be systematically derived \blue{and revealed a discontinuous transition from poor to perfec generalization}~\cite{Gyo-1990,Sompolinsky-1990}. In contrast to the classical bias-variance trade-off (U-shaped curve of the test error versus increasing model complexity)~\cite{Mehta-2019}, the modern
deep learning achieves the state-of-the-art performance in the over-parameterized regime~\cite{Belkin-2019,Spigler-2019}, \blue{a regime of the number of parameters much larger than the training data size.}
However, how to provide an analytic argument about the over-fitting effects versus different
parameterization regimes (e.g., under-, over- and even super-parameterization) for this empirical observation
becomes a non-trivial task~\cite{Jeff-2020}.  A recent study of one-hidden-layer networks shows that a first transition occurs at the interpolation point, where perfect fitting becomes possible. This transition reflects the properties of hard-to-sample typical solutions. Increasing the model complexity, a second transition occurs with the discontinuous appearance of atypical solutions. They are wide minima of good generalization properties. This second transition sets an upper bound for the effectiveness of learning algorithms~\cite{Baldassi-2022}. \blue{This statistical mechanics analysis focuses on the average case (average over all realizations of data, network and algorithm), rather than the worst case. The worst case determines the computational complexity category, while the average case tells us the universal properties of learning, and the statistical mechanics links the computational hardness to a few order parameters in physics~\cite{HH-2022}, and these previous works show strong evidences~\cite{Huang-2014pre,Baldassi-2016,Spigler-2019,Huang-2020,Baldassi-2022}.}

 For an infinitely wide neural network, there exists a lazy learning regime, where the overparameterized neural networks can be well approximated by a linear model corresponding to a first-order Taylor expansion around the initialization, and the the complex learning dynamics is simply training a kernel machine~\cite{Belkin-2021}. \blue{However, in a practical training, the dynamics is prone to escape the lazy regime, which has no satisfied theory yet}. Therefore,
clarifying which of lazy-learning (or neural tangent kernel limit) and feature-learning (or mean-field limit) may explain the success of deep supervised learning remains open and challenging~\cite{Jacot-2018,Zhang-2021,Monta-2021}. \blue{The mean-field limit can be studied in the field theoretic framework, characterizing how the solution of learning deviates from the initialization through a systematic perturbation of the action in the framework~\cite{Helias-2022}. Another related challenge is out-of-distribution generalization, which can also be studied using statistical mechanics, e.g., in a recent work, a kernel regression was analyzed~\cite{OOD-2021}. In addition, the field theoretic method is also promising to write the learning problem of out-of-distribution prediction into propagating correlations and responses~\cite{Helias-2022}.} 
 
\section{ Challenge \Rmnum{3}---  Adversarial vulnerability} 
Adversarial examples are defined by those inputs with human-imperceptible modifications yet leading to unexpected errors 
in a deep learning decision making system. \blue{The test accuracy drops as the perturbation grows; the perturbation can either rely on the trained network or be an independent noise~\cite{Szegedy-2014,Good-2015,ZJ-2020}. The current deep learning is argued to learn predictive yet non-robust features in the data~\cite{Shortcut-2020}.} This adversarial vulnerability of deep neural networks poses a significant challenge in the practical applications of both 
real-world problems and AI4S (artificial intelligence for science) studies. \blue{Adversarial training remains the most effective solution to the problem~\cite{Madry-2018}, yet in contrast to human learning. However, the training sacrifices the standard discrimination. A recent work applied the physics principle that the hidden representation is clustered like replica symmetry breaking in spin glass theory~\cite{Mezard-1987}, which leads to contrastive learning that is local and adversarial robust, resolving the trade-off between standard accuracy and adversarial robustness~\cite{Xie-2024}. Furthermore, the adversarial robustness can be theoretically explained in terms of a cluster separation distance. } 
 In physics, systems with a huge number of degrees of freedom are able to be captured by a low-dimensional macroscopic description, such as Ising ferromagnetic model. Explaining the layered computation in terms of geometry may finally help to crack the mysterious property of the networks' susceptibility to adversarial examples~\cite{Fellow-2018,Borto-2018,Li-2023}.
Although some recent efforts were devoted to this direction~\cite{Kenway-2018,Borto-2018}, more exciting results are expected in near future works.

\section{Challenge \Rmnum{4}---  Continual learning}
 A biological brain is good at adapting the acquired knowledge from similar tasks to domains of new tasks, even if only a handful examples
are available in the new-task domain. \blue{This kind of learning is called continual learning or multi-task learning~\cite{CF-1989,Kirk-2017}, an ability to learn many tasks in sequence, while transfer learning refers to the process of 
exploiting the previously acquired knowledge from a source task to improve the generalization performance in a target task~\cite{CLL-2019}.}  However, the stable adaptation to changing environments, an essence of lifelong learning, remains a significant challenge for modern artificial intelligence~\cite{CLL-2019}. More precisely, neural networks are in general poor at the multi-task learning, although impressive progresses have been achieved in 
recent years. For example, during learning, a diagonal Fisher information term is computed to measure importances of weights (then a rapid change is not allowed for those important weights) for previous tasks~\cite{Kirk-2017}.
A later refinement by allowing synapses accumulating task relevant information over time was also proposed~\cite{Zenke-2017}.
More machine learning techniques to reduce the catastrophic forgetting effects were summarized in the review~\cite{CLL-2019}. However, 
we still do not know the exact mechanisms for mitigating the catastrophic forgetting effects in a principled way, which calls for theoretical studies of 
deep learning in terms of adaptation to domain-shift training, i.e., connection weights trained in a solution to one task are transformed to benefit learning on a related task.

\blue{Using asymptotic analysis, a recent work studying transfer learning identified a phase transition in the quality of the knowledge transfer~\cite{Lu-2021}.
This work reveals how the related knowledge contained in a source task can be effectively transferred to boost the performance in a target task.}
 Other recent theoretical studies interpreted the continual learning with a statistical mechanics framework using Franz-Parisi potential~\cite{Huang-2023} \blue{or as an on-line mean-field dynamics of weight updates~\cite{ICML-2021}}. \blue{The Franz-Parisi potential is a thermodynamic potential used to study glass transition~\cite{Franz-1995}. The recent work assumes that the knowledge from the previous task behaves as a reference configuration~\cite{Huang-2023}, where the previously acquired knowledge serves as an anchor for learning new knowledges.} This framework also connects to elastic weight consolidation~\cite{Kirk-2017}, heuristic weight-uncertainty modulation~\cite{VCL-2020}, and neuroscience inspired metaplasticity~\cite{meta-2021}, providing a theory-grounded method for the real-world multi-task learning with deep networks.

\section{Challenge \Rmnum{5}---Causal learning}
Deep learning is criticized as being nothing but a fancy curve-fitting tool, making a naive association between inputs and outputs.
In other words, this tool could not distinguish correlation from causation. What the deep network learns is not a concept but merely a statistical correlation, prohibiting the network from counterfactual inference (a hallmark ability of intelligence).
A human-like AI must be good at retrieving causal relationship among feature components in sensory inputs, thereby carving relevant information from a sea of
irrelevant noise~\cite{Kopf-2019,Pearl-2018,CL-2021}. Therefore, understanding cause and effect in deep learning systems is particularly important for the next-generation artificial intelligence.
The question whether the current deep learning algorithm is able to do causal reasoning remains open. Hence, \blue{how to design a learning system that can infer the effect of an intervention becomes a key to 
address this question, although it would be very challenging to make deep learning extract causal structure behind observations by applying simple physics principles, due to both architecture and learning complexities}. This challenge is now intimately related to the astonishing performances of large language models (see the following seventh challenge), \blue{and the key question is whether the self-attention mechanism is sufficient for capturing the causal relationships in the training data}.

\section{Challenge \Rmnum{6}---Internal model of the brain} 
The brain is argued to learn to build an internal model of the outside world, reflected by 
spontaneous neural activities as a reservoir for computing (e.g., sampling)~\cite{Dario-2009}. The agreement between spontaneous activity and stimulus-evoked
one increases during development especially for natural stimuli~\cite{Berkes-2011}, while the spontaneous activity outlines the regime of evoked neural responses~\cite{Luczak-2009}. \blue{The relationship between the spontaneous fluctuation and task-evoked response causes recent interests in studying brain dynamics~\cite{Deco-2023}. This can be formulated by the fluctuation-dissipation theorem in physics, and the violation can be a measure of deviation from equilibrium, although a non-equilibrium stationary state exists.}

In addition, the stimuli were shown to carve a clustered neural space~\cite{Huang-2016a,Berry-2020}. Then, an interesting question is what the spontaneous neural space looks like, and how the space dynamically evolves, especially in adaptation to changing environments. Furthermore, 
how sensory inputs combined with the ongoing asynchronous cortical activity determine the animal behavior remains open and challenging.
If the reward mediated learning is considered, reinforcement learning was used to build world models of structured environments~\cite{Ha-2018}. In the reinforcement learning, observations are used to 
drive actions which are evaluated based on reward signals the agent receives from the environment after taking the actions. It is thus interesting to reveal which kind of internal models the agent establishes through learning from interactions with
the environments. This can be connected to aforementioned representation and generalization challenges.
Moreover, a recent work showed a connection between the reinforcement learning and statistical physics~\cite{RLSP-2019},  suggesting that a statistical mechanics theory could be potentially established to understand \blue{how an optimal policy is found to maximize the long-term accumulated reward}, with an additionally potential impact on studying reward-based neural computations in the brain~\cite{RL-2019}.

Another angle to look at the internal model of the brain is through the lens of neural dynamics~\cite{Deco-2009,Sussillo-2009,Buonomano-2009,Sussillo-2020}, which is placed onto a low dimensional surface, robust to variations in detailed properties of individual neurons or circuits. The representation of stimuli, tasks or contexts can be retrieved for deriving experimentally testable hypotheses~\cite{Ostojic-2021}. \blue{Although previous theoretical studies were carried out in recurrent rate or spiking activity neural networks~\cite{Sompolinsky-1988,Brunel-2000}}, a challenging issue remains to address how neural activity and synaptic plasticity interact with each other to yield a low dimensional internal representation for cognitive functions. The recent development of synaptic plasticity combining connection probability, local synaptic noise and neural activity can realize a dynamic network in adaptation to time-dependent inputs~\cite{Zou-2023}. This work interprets the learning as a variational inference problem, making optimal learning under uncertainty possible in a local circuit. Both learning and neural activity are placed on low-dimensional subspaces. Future works must include more biological plausible factors to test the hypothesis in neurophysiological experiments. \blue{Another recent exciting achievement is using dynamical mean-field theory to uncover rich dynamical regimes of coupled neuronal-synaptic dynamics~\cite{Clark-2024}.}

Brain states can be considered as an ensemble of dynamical attractors~\cite{Von-2018}. The key challenge is how learning shapes the stable attractor landscape. One can interpret the learning as a Bayesian inference, e.g., in an unsupervised way, but not the autoregressive manner (see the next section). The learning can then be driven by synaptic weight symmetry breaking~\cite{Hou-2019,Huang-2020}, separating two phases of recognizing the network itself and the rule hidden in sensory inputs. It is very interesting to see if this picture still holds in recurrent learning supporting neural trajectories on dynamical attractors, \blue{and even predictive learning minimizing a free energy of belief and synaptic weights (the belief leads to error neurons)~\cite{Rao-2024}. New methods must be developed, e.g., based on recently proposed quasi-potential method to study non-equilibrium steady neural dynamics~\cite{Qiu-2024}, or dynamical mean-field theory for learning~\cite{DMFT-2023}.}

\section{Challenge \Rmnum{7}---Large language models}
The impressive problem-solving capabilities of Chat-GPT where GPT is a shorthand of generative pretrained transformer are leading the fourth industrial revolution. The Chat-GPT is based on large language models (LLMs)~\cite{GPT-4}, which represents linguistic information as vectors in high dimensional state space, trained with a large text corpus in an autoregressive way (in analogy to the hypothesis that the brain is a prediction machine~\cite{Clark-2013}), resulting in a complex statistical model of how the tokens in the training data correlate~\cite{Transformer-2017}. The computational model thus shows strong formal linguistic competence~\cite{Tenen-2023}.
The LLM is also a few-shot or even zero-shot learner~\cite{LLM-2020,LLM-2022}, i.e., the language model can perform a wide range of computationally challenging tasks with prompting alone (e..g, chain-of-thought prompting~\cite{COT-2022}). Remarkably,
the LLMs display a qualitative leap in capability as the model complexity and sample complexity are both scaled up~\cite{Kaplan-2020}, akin to phase transitions in thermodynamic systems.

 In contrast to the formal linguistic competence, the functional linguistic competence is argued to be weak~\cite{Tenen-2023}. This raises a fundamental question what the nature of intelligence is, or whether a single next-token context conditional prediction is a standard model of artificial general intelligence~\cite{Sej-2023,Lake-2017,Gerven-2017}. Human's reasoning capabilities in real-world problems rely on non-linguistic information as well, e.g., it is unpredictable when a creative idea for a scientist would come to a challenging problem at hand, which relies on reasoning about the implications along a sequence of thought. In a biological implementation, the language modules are separated from the other modules involving high-level cognition~\cite{Tenen-2023}. The LLM
explains hierarchical correlations in word pieces in the training corpora, rather than hidden casual dependencies. In other words, the neural network has not constructed a mental model of the world, which requires heterogeneous modular networks, thereby unlike humans. Therefore, the LLM does not know what it generate (as a generative model). Even if some key patterns of statistical regularities are absent in the training data, the model can generate perfect texts in terms of syntax. However, the texts may be far away from the truth. Knowing what they know is a crucial hallmark of intelligent systems~\cite{Gerven-2017}. In this sense, the inner workings of the LLM are largely opaque, requiring a great effort to mathematically formulate the formal linguistic competence, and further identify key elements that must be included to develop a robust model of the world. Mechanisms behind the currently observed false positive like hallucination~\cite{Chom-2023} could then be revealed, \blue{which may be related to interpolation between modes of token distributions. A recent work interpreting the attention used in transformer-based LLM as a generalized Potts model in physics seems inspiring~\cite{Potts-2024}, i.e., tokens as Potts spin vectors.}

Most importantly, we currently do not have any knowledge about how to build an additional network that is able to connect performance and awareness~\cite{Axel-2014}, which is linked to what makes us conscious (see the last challenge). Following the Marr's framework, both computational and neural correlates of consciousness remain unknown~\cite{Crick-2003,PNAS-2022CS,NCC-2023}.
A current physical way is to consider a Lyapunov function governing complex neural computation underlying LLMs~\cite{Hopfield-2020,Krotov-2020}.
In this way, the Lyapunov function perspective will open the door of many degrees of freedom to control how information is distilled via not only the self-attention but also other potential gating mechanisms, \blue{based on dynamical system theory.}

\section{Challenge \Rmnum{8}---Theory of consciousness} 

One of the most controversial questions is the origin of consciousness---whether the consciousness is an emergent behavior of highly heterogeneous and modular
brain circuits with various carefully-designed regions (e.g., a total of about $10^{14}$ connections for human brain and many functionally specific modular structures, such as Prefrontal cortex, Hippocampus, Cerebellum, etc~\cite{Harris-2015,Luo-2021}). The subjectivity of the conscious experience is in contradiction with 
the objectivity of a scientific explanation. According to the Damasio's model~\cite{Self-2001}, the ability to identify one's self in the world and its relationship with the world is considered to be a
central characteristic of conscious state. Whether a machine algorithm can achieve the self-awareness remains elusive. The self-monitoring ability (or meta-cognition~\cite{Deha-2017}) may endow the machines (such as LLMs) to know what they generate. It may be important to clarify how the model of self is related to the internal model of the brain (e.g., through recurrent or predictive processing~\cite{Senn-2024}). For example, Karl Friston argued that the conscious processing can be interpreted as a statistical inference problem of inferring causes of sensory observations. Therefore, minimizing the surprise (negative log probability of an event) may lead to self-consciousness~\cite{Friston-2018}, in consistent with the hypothesis that the brain is a prediction machine~\cite{Clark-2013,Gerven-2017}.

There are currently two major cognitive theory of consciousness. One is the 
global workspace framework~\cite{Dehaene-1998}, which relates consciousness to the widespread and sustained propagation of cortical neural activities by demonstrating that 
consciousness arises from \blue{an ignition that leads to global information broadcast among brain regions. This computational functionalism was recently
 leveraged to discuss possibility of consciousness in non-organic artificial systems~\cite{Bengio-2017,Bengio-2023}.}  The other is the integrated information theory that provides a quantitative characterization of conscious state 
by integrated information~\cite{Tononi-2004}. In this second theory, unconscious states have a low information content, while conscious states bear a high information content. \blue{The second theory emphasizes the phenomenal properties of consciousness~\cite{IIT4p0}, i.e., the function performed by the brain is not subjective experience.}
 Both theories follow a top-down approach, which is in stark contrast to the statistical mechanics approach following a bottom-up manner building the bridge from 
microscopic interactions to macroscopic behavior. These hypotheses are still under intensive criticism despite some cognitive experiments they can explain~\cite{Koch-2016}. We remark that conscious states may be an emergent property of neural activities, lying at a higher level than neural activities. It is currently unknown how to connect these two levels, for which a new statistical mechanics theory is required. \blue{An exciting route is to link the spontaneous fluctuation to stimulus-evoked response, and a maximal response is revealed in a recurrent computational model~\cite{Qiu-2024}, which can be thought of as a necessary condition for consciousness, as information-richness of cortical electrodynamics was also observed to be peaked at the edge-of-chaos (dynamics marginal stability)~\cite{Toker-2022}. This peak thus distinguishes the conscious from unconscious brain states.}
From an information-theoretic argument, the conscious state may require a diverse range of configurations of interactions between brain networks, which can be linked to the entropy concept in physics~\cite{Erra-2016}.
The large entropy leads to optimal segregation and integration of information~\cite{Zhou-2015}. 

Taken together, whether the consciousness can be created from an interaction of local dynamics within complex neural substrate is still unsolved~\cite{Krauss-2020}. A statistical mechanics theory, if possible, is always promising in the sense 
that one can make theoretical predictions from just a few physics parameters~\cite{HH-2022}, which may be possible from a high degree of abstraction and thus a universal principle could be expected.

\section{Conclusion}
To sum up, in this viewpoint, we provide some naive thoughts about fundamental important questions related to neural networks, for which building a good theory is far from being completed. The traditional researches of statistical physics of neural networks bifurcate to two main streams: one is to the engineering side, developing theory-grounded algorithms; and the other is to the neuroscience side, formulating brain computation by mathematical models solved by physics methods. 
In physics, we have the principle of least action, from which we can deduce the classical mechanics or electrodynamics laws. We are not sure whether in physics of neural networks (and even the brain) there exists
general principles that can be expressed in a concise form of mathematics.
\blue{It is exciting yet challenging to promote the interplay between physics theory and neural computations along these eight open problems discussed in this perspective paper. The advances will undoubtedly lead to a human-interpretable understanding of underlying mechanisms of the artificial intelligent systems, the brain and mind, especially in the era of big experimental data in brain science and rapid progress in AI researches. }

\begin{acknowledgments}
We would like to thank all PMI members for discussions lasting for five years. This perspective also benefits from discussions with students during the on-line course of statistical mechanics of neural networks (from September 2022 to June 2023).  We are also grateful to invited speakers in the INTheory on-line seminar during the COVID-19 pandemic. We enjoyed a lot of interesting discussions with Adriano Barra, Yan Fyodorov, Sebastian Goldt, Pulin Gong, Moritz Helias, Kamesh Krishnamurthy, Yi Ma, Alexander van Meegen,  Remi Monasson, Srdjan Ostojic, Riccardo Zecchina.
This research was supported by the National Natural Science Foundation of China for
Grant numbers 12122515. 
\end{acknowledgments}


\end{document}